\documentclass[aps,prE,twocolumn,showpacs,showkeys,groupedaddress]{revtex4}% for review and submission
\usepackage{graphicx}  % needed for figures
\usepackage{dcolumn}   % needed for some tables
\usepackage{bm}        % for math
\usepackage{amssymb}   % for math
\usepackage{subfigure}
\usepackage{amsmath,amsfonts}
\usepackage{setspace}
\usepackage[mathscr]{euscript}
\usepackage[utf8]{inputenc}

\begin{document}
\title{Topological phase transitions in 2-dimensional bent-core 
liquid crystal models}

\author{B. Kamala Latha$^{1}$}
\author{Surajit Dhara$^{1}$}
\author{V.S.S. Sastry$^{2}$}
\affiliation {$^{1}$School of Physics,University of Hyderabad, Hyderabad 500046, India}

\affiliation{$^{2}$Centre for Modelling, Simulation and Design, University of Hyderabad, Hyderabad 500046, India}

\date{\today}% 

\begin{abstract}

Spontaneous onset of a low temperature topologically ordered phase 
in a 2-dimensional (2D) lattice model of uniaxial liquid crystal (LC)  
(\textit{d} = 2, \textit{n} = 3), was debated extensively in view of earlier  
Monte Carlo (MC) results pointing to a suspected underlying mechanism 
affecting the RG flow near the topological fixed point. A recent MC 
study clarified that a prior crossover leads to a transition to  nematic 
phase. The crossover was interpreted as due to the onset of a 
perturbing  relevant scaling field originating from the extra spin 
degree of freedom. As a counter example and in support of this 
hypothesis, we have been investigating 2D lattice models of the 
LC system, which include 
all three $\textit{'spin'}$ degrees in the interaction terms of the 
Hamiltonian. In this work, we consider V-shaped bent-core molecules 
with rigid rod-like segments connected at an assigned angle. The 
two segments of the molecule interact with the segments of all the 
nearest neighbours on a square lattice, prescribed by a 
biquadratic interaction. We compute equilibrium averages of 
different observables with Monte Carlo techniques (employing 
Metropolis sampling as well as entropic sampling algorithms, 
for comparison), as a function of temperature ($T$) and sample size. 
Besides the system energy, the specific heat and the orientational 
order parameters of the medium, we computed additional variables 
that track distinct signatures of potential topological defect-mediated 
Berezinskii-Kosterlitz-Thouless (BKT) type transitions (topological order 
parameters and densities of unbound topological defects, of the 
different ordering directors). For the chosen molecular bend angle 
($\theta = 112^{\circ}$)  and symmetric inter-segment interaction between neighbouirng molecules, the 2D system shows two transitions 
as a function of $T$: the higher one (at $T_{1}$) leads to a topological 
ordering of defects associated with  the major molecular axis 
(say z-axis), without a crossover, imparting uniaxial symmetry to the 
medium described by the first fundamental group of the order parameter 
space $\pi_{1}$= $Z_{2}$  (inversion symmetry). The 
second at $T_{2}$ leads to a medium displaying biaxial symmetry with 
$\pi_{1}$ = Q (quaternion group).  The correlation functions of 
the three molecular axes $G(r,T)$ corroborate these observations. 
The high temperature phase shows single length scale
to each of  the three axes (exponential decays of $G(r)$) except near the 
pre-transitional region. The biaxial phase shows a self-similar 
microscopic structure with the three axes  showing power law 
correlations with vanishing exponents as the temperature decreases. 
The qualitative differences in the topological parameters of the 
minor axes as well as in their correlation functions, show their 
asymmetric coupling to the z-axes. This appears to be associated 
with the specific local site symmetry of the molecular V-shape 
geometry, being distinct from the global symmetry of the order 
parameter space in the biaxial phase. In terms 
of a phenomenological parameterization of this model, the results 
indicate a significant degree of cross coupling between the 
uniaxial and biaxial basis tensors of the interacting molecules.

\end{abstract}
\maketitle
\section{Introduction}
\label{}
The self organizing property of soft matter leads to myriad applications 
in the fields of optics, biosensors and electronics. The formation of
 liquid crystal (LC) phases on 2-dimensional (2D) surfaces is the key to 
 many nanotechnological applications \cite{Dierking, Gong}. Phases 
 with conventional long range order cannot exist in continuous systems 
 with short range interactions in lower physical dimensions $d\leq 2$
 \cite{Mermin}. However low temperature phases with quasi-long range order
 (QLRO) are still realizable in such systems if stable topological 
 defects are permitted in the system  due to non-trivial topology of 
 the order parameter space ($\mathcal{R}$), - a phenomenon first detected 
 in the 2D (\textit{d}=2) XY magnetic model through the Berezinskii-Kosterlitz-Thouless (BKT) mechanism \cite{Berz2, Kosterlitz}.  Liquid crystals 
 with global SO(3) symmetry and a local site symmetry of $Z_{2}$ 
 (uniaxial systems) have an order parameter (OP) space which is not 
 simply connected. This OP geometry leads to the presence of 
 point defects in two dimensional uniaxial as well as biaxial nematics. 
 The fundamental group of the OP space for uniaxial nematics (with 
 $D_{\infty h}$ point group symmetry) is isomorphic to the two element 
 Abelian group $\Pi_{1}(\mathcal{R})  = Z_{2}$.  In two dimensions 
 this provides  for stable point disclinations with topological 
 charge (winding number) with a value ${\pm}$ 1/2 , corresponding 
 to rotation of the order director by  $180 ^{\circ}$ (in OP space) 
 for a closed path in the physical space. For biaxial nematics which 
 have a global group point symmetry $D_{2}$, the order parameter space is 
 $\mathcal{R} = SO(3)/D_{2}$. Taking the homomorphic correspondance between 
 SO(3)  and SU(2), this  becomes $\mathcal{R}= SU(2)/Q$ where $Q$ is the lift 
 of $D_{2}$ in SU(2). Thus the fundamental group of biaxial nematics 
 is $\Pi_{1}(\mathcal{R})= Q$, the eight element quarternion group 
 which is discrete and non-Abelian. The eight
  elements can be grouped into five conjugacy classes $C_{0}$, 
  $\overline{C}_{0}$, $C_{x}$, $C_{y}$ and $C_{z}$. The class 
  $C_{0}= [1]$ contains  removable defects, 
 $\overline{C}_{0}=[-1]$ contains $360^{\circ}$ disclinations  and 
 classes $C_{x} = [\pm i\sigma_{x}]$, $C_{y} = [\pm i\sigma_{y}]$ and 
 $C_{z}=[\pm i\sigma_{z}]$ (where $\sigma_{x}$, $\sigma_{y}$, $\sigma_{z}$ 
 are Pauli  matrices) contain defects in which the rotation 
 is through  ${\pm}$ $180^{\circ}$ about each of the distinct symmetry 
 axes. Thus stable defects of both integer and half integer charges 
 exist in  biaxial nematics \cite{Mermin}.  
 
The BKT disclination unbinding scenario in the 
(\textit{n}=2, \textit{d}=2) uniaxial lattice model (Lebwohl-Lasher
interaction), wherein the reorientations of moelcules on the 2d-lattice 
sites (\textit{d}=2) are restricted to a plane (\textit{n}=2), is 
formally equivalent to the 2D-XY model, as was  verified through 
simulations. This model exhibits a low temperature nematic phase
(of uniaxial symmetry) with a QLRO phase.  Similar observations 
were made in 2D simulations of hard rods \cite{Eppenga, Vink}, 
spherocylinders \cite{Bates}. Recent simulations of
hard bent  needles \cite{Tavarone} in 2D (\textit{n}=2,\textit{d}=2)
have also shown evidence of disclination unbinding leading
to isotropic-quasi nematic phase transition. 

In this paper, we investigate the phase behaviour of a bent-core lattice 
model in two-dimensions (\textit{d}=2, \textit{n}=3) wherein the molecules
having spin dimensionality (\textit{n}=3) are confined to a 
plane (\textit{d}=2). This study is motivated by a recent investigation of 
a general biquadratic Hamiltonian \cite{Sonnet, Bisi,
 Romano} with a $D_{2h}$ site  symmetry (\textit{d}=2, \textit{n}=3)  
 \cite{BKL20}, with certain simplifying conditions. It is assumed 
 in \cite{BKL20}  that in the Hamiltonian comprising of general
 biquadratic interactions among the molecules, the interaction of the 
 uniaxial tensor (associated with molecular major axis of any molecule) 
 with the biaxial tensors (of the minor axes of the interacting 
 molecules) is not present, and the model parameters are chosen to 
 induce a direct strong transition from the disordered state to a 
 biaxial low temperature phase. The molecular site is assigned symmetry 
 identical to that of the biaxial medium. A detailed Monte Carlo study
  demonstrated a topological transition to a low temperature phase 
  with $D_{2h}$  global symmetry, characterized by a line of critical 
  points. These observations are supported by the computed topological 
  variables and their low temperature limits. This work, besides 
  confirming a BKT-type topological transition in a 2D biaxial 
  phenomenological model, also provided independent evidence, 
 supporting the conjecture made to explain the observed crossover in  
 the 2D uniaxial model \cite{BKLPRL}.  The addition of a 
  suitable biaxial perturbation to the 2D uniaxal model  
 explicitly pointed to the necessity to reduce the topological 
 symmetry of the OP space so as to make its fundamental group discrete and  
 non-Abelian, making it a necessary criterion for this interesting defect 
 mediated transition \cite{BKL20}. In this context, the current study examines 
 this hypothesis by choosing a biaxial lattice model (bent-core type)
 with a local site symmetry different from the global symmetry of the 
 low temperature phase.  It is based on a microscopic molecular level 
 prescription of interactions, which facilitates explicit choices 
 regarding the molecular geometry and the degree of different interactions.  
  
To provide an appreciation of the chosen bent-core model within 
the context of the well-established  phenomenological studies, it is useful 
to briefly introduce the general biquadratic Hamiltonian for biaxial 
systems accounting for the interactions among the molecular tensors 
with uniaxial and biaxial symmetry, with three phenomenological model 
parameters. Its phase diagram in three dimensions was extensively 
investigated earlier \cite{Sonnet, Bisi, Romano, BKL18}, specifically 
providing insights into the role of the two types of biaxial coupling 
hosted in the Hamiltonian. The recent 2D-biaxial  study mentioned 
above \cite{BKL20} is a particularly simplifying choice of these parameters.      

The interaction between two lattice sites, each possessing in general
 a biaxial symmetry, is expressed in terms of the two orthogonal and 
 traceless molecular tensors associated with each site:
$\bm{q} := \bm{m} \otimes \bm{m} - \frac{\bm{I}}{3}$ 
(uniaxial symmetry) and
$\bm{b} := \bm{e} \otimes \bm{e} - \bm{e}_{\perp} \otimes \bm{e}_{\perp}$
(biaxial symmetry). Here $(\bm{e},\bm{e_{\perp}},\bm{m})$ is an 
orthonormal set of vectors representing the molecular axes 
(in the notation of \cite{Sonnet}). System Hamiltonian, inclusive 
of the biaxial symmetry, is expressed as a general interaction 
between two lattice sites (i, j) :  
$ H=-U [\xi \, \bm{q_{i}} \cdot \bm{q_{j}}
+ \gamma(\bm{q_{i}} \cdot \bm{b_{j}} + \bm{q_{j}}
 \cdot \bm{b_{i}}) + \lambda \, \bm{b_{i}} \cdot \bm{b_{j}}]$.
The Lebwohl-Lasher (LL) model corresponds to limiting H to 
uniaxial symmetry, by setting $\xi = 1$, $\gamma = 0$, 
$\lambda = 0$.  The 2D-LL model (with \textit{d}=2, \textit{n} =3)
was extensively studied  both with Metropolis algorithm 
\cite{Kunz, Shabnam} and recently with entropic sampling 
method \cite{BKLPRL}. The phenomenological 2D biaxial
 model referred to earlier \cite{BKL20} corresponds to setting 
 $\xi = 1$, $\gamma = 0$, $\lambda = 1/3$. It is expected that the 
 bent-core model could be a promising candidate to provide an insight 
 into the phase behaviour of a  general model with contributions from 
 all the three terms. With the added flexibility as mentioned, and 
 maintaining the site symmetry always different from the global symmetry,
 this model facilitates effective variation of the parameters 
 $\gamma$ and $\lambda$ in the phenomenological model.

 The paper is organised as follows. In section II we present the 
 Hamiltonian Model and the  simulation details. The data are 
 presented and results of their detailed analysis are discussed 
 section III. We conclude with a summary of the salient features 
 of this work in section IV.  
 \section{Model and Simulation details}
\label{•}
\subsection{Model and Hamiltonian}

 We investigated the two dimensional phase behaviour
 of symmetric V-shaped bent core molecules bent at an interarm 
 angle of $\theta$ and with equal interarm coupling strengths
set to unity. The symmetry axes  for these molecules are:
(i) the axis orthogonal to the molecular plane (x-axis); 
(ii) the axis bisecting the interarm angle (y-axis); and (iii) the 
axis mutually perpendicular to these two axes (z-axis).
When the arms are orthogonal to each other 
($\theta = 90^{\circ}$) the interaction tensor for these molecules
is cylindrically symmetric about the axis perpendicular to the arms.
The molecule is disk-like for angles 
 $90^{\circ} < \theta < 109.47^{\circ}$ and rod-like for angles
 $109.47^{\circ} < \theta < 180^{\circ}$, $\theta = 109.47^{\circ}$
 being the angle of tetrahedral geometry for the molecule. As the 
 interarm angle increases from $90^{\circ}$ to $180^{\circ}$
the molecular interaction tensor can be considered to have 
contributions from both uniaxial and biaxial tensorial components, the 
relative importance of which determine the formation of the 
respective phases. The simulation studies of phase
behaviour of these molecules in three dimensions (\textit{d}=3) 
interacting through an attractive potential  has revealed an 
Isotropic (I)-Uniaxial nemtatic ($N_{U}$)-Biaxial nematic ($N_{B}$)
phase sequence for various interarm angles and different interaction
strengths between the arms. A direct Isotropic - Biaxial
transition is also predicted for the interarm angle of 
$\theta = 109.47^{\circ}$, for unit interaction strength
 \cite{Bates2005}. In this study, we chose a suitable value of 
$\theta$ such that a prolate uniaxial phase is formed 
on condensation from the isotropic phase by alignment of the
long molecular axes (z-axes). On further condensation, a biaxial phase 
forms by the alignment of the short molecular axes(x- and y-axes).  
Due to the non-trivial topological symmetry of the OP space of this
model, the fundamental group $\Pi_{1}(R)= Q$, and hence stable 
topological defects with differing charges (winding numbers) exist 
in the medium.  We expect, from earlier two-dimensional work, this 
model to host thermally induced topological transitions, with an 
underlying BKT-type unbinding mechanism primarily involving half-charge 
disclinations associated with the three director axes. The topologically 
ordered phases, if the medium condenses to, are evidenced by their 
characteristic microscopic structures with quasi long-range order 
(QLRO), and sharp variations of the topological variables which 
quantify the degree of binding of the point defects.
  
The V-shaped LC molecules are modelled as simple extension of the 
uniaxial LL lattice model, suggested by Bates \textit{et al} 
\cite{Bates2005}. Here, a lattice site hosts a mesogenic molecule 
made up of two rod like  constituents A and B joined at a fixed 
angle $\theta$. These two constituents interact with those of four 
nearest neighbour molecules on the square lattice. The potential 
between two neighbours (with identical constituents) at two such lattice 
sites \textit{i} and \textit{j} is expressed as: 
%\begin{widetext}
\begin{equation}
U(\omega_{ij}) =- \sum_{\alpha=A,B} \ \sum_{\beta=A,B}
 \epsilon_{\alpha\beta} \ P_{2}(\cos (\gamma_{\alpha\beta})
\label{eqn:1}
\end{equation}
%\end{widetext}
where the indices $\alpha$ and $\beta$ run over the two constituent 
segments of each molecule on the sites at \textit{i} and \textit{j}, 
respectively. The angle $\gamma_{\alpha\beta}$ 
is between the rod $\alpha$ in molecule \textit{i} and rod $\beta$ in 
molecule \textit{j}. For symmetric molecules, the anisotropy of interaction 
for each arm is the same and is given by 
$\epsilon_{AA}=\epsilon_{AB}=\epsilon_{BB}= \epsilon_{BA}$. 
$\epsilon_{AA}$ is used as a scale factor for the  potential. 
The reduced temperature scale for the simulation is set as 
$T = \frac{K_{B}T^{'}}{\epsilon_{AA}}$ where
$T^{'}$ is the laboratory temperature (in Kelvin).

 \subsection{Simulation Details}
 Simulations were carried out using both the Metropolis based Monte 
 Carlo (MC) method and entropic sampling method guided by a modified 
 Wang-Landau algorithm. The Metropolis algorithm \cite{Metropolis} 
 based on the Markov chain Monte Carlo sampling facilitates an 
 otherwise perfectly random walk of the system in the (ergodic) 
 configuration space, but for guidance for the acceptance or otherwise 
 of each random  step respecting Boltzmann equilibration criterion at 
 the chosen temperature $T$. The underlying algorithm ensures that the 
 system, starting from an arbitrary initial state, converges to a 
 sequence of equilibrated microstates in  the asymptotic limit of a 
 long enough walk. Large enough set of equilibrated states follows the 
 canonical distribution at $T$, constituting the Boltzmann ensemble (B-ensemble). 
 The averages of relevant physical observables are computed as averages
 over these microstates.   
             
Wang-Landau sampling procedure \cite{Wang} on the other hand guides 
the system asymptotically to perform a random walk which is uniform 
with respect to system energy, estimating in the process the 
representative density of states (DoS) of the system  $g(E)$ 
with respect to its energy. This algorithm is now generalized to 
be applicable in different areas of research, like finite density 
quantum field theories \cite{Langfeld}, complex magnetic systems
\cite{Brown}, polymers and protein folding \cite{Shi}, and spin 
cross-over systems \cite{Chan}and  is being continually updated 
for parallel processing on multiple nodes using, for example, 
replica exchange protocol \cite{Vogel}. The algorithm  has been 
modified for different model systems, for example for lattice models 
such as Lebwohl-Lasher interaction \cite{LL} which requires continuous 
molecular reorientations in liquid crystal models \cite{Jayasri}. It 
was further augmented by the so-called frontier sampling technique 
\cite{Zhou,BKL15}. It is an algorithm to force the system to visit 
progressively lower energy states (with  extremely low probability), by 
setting up energy barriers at chosen points on the energy axis (referred 
to as frontiers), thereby discouraging access to higher energy regions where 
the DoS has been already estimated approximately. This process is continued 
till the desired energy range is covered. The underlying guiding 
distribution function generated by continuous upgrades during this walk 
provides an approximate estimate of $g(E) $. At this stage the random 
walk is allowed to proceed according to the Wang-Landau method, but 
without further insertion of  energy barriers. The updating of the 
distribution function is continued while gradually reducing the 
algorithmic guidance, till $g(E) $ is determined to the desired 
accuracy, normally limited by the computational accuracy. This limit of 
 $g(E) $ is the representative DoS of the system. A large entropic 
 ensemble of microstates  ( $\sim 10^{8}$) is then collected by 
 performing a random walk 
 in the configuration space with an acceptance probability based on the 
 inverse of $g(E)$. The entropic ensemble for a well estimated $g(E)$ 
 is reasonably uniformly distributed with energy (despite huge change 
 in the entropies), typically to within 15-20$\%$. The relevant 
 equilibrium averages of observables are computed at a desired 
 temperature by deriving equilibrium ensembles (RW-ensembles)
  from the set of states in the entropic ensemble, by the standard 
  reweighting procedures \cite{Swensden, Berg}.  Further details of this 
  modified  Wang-Landau  algorithm augmented by frontier sampling 
  can be found in  Ref \cite{BKL15, BKL18}. Besides the richness of 
  the configuration space in terms of accessible states in such 
  LC models due to continuos  random steps of the system, another 
  factor which significantly puts a huge demand on the computational 
  time is the shape of the molecule and the prescription of the 
  Hamiltonian. In the present model, the computational 
 effort is considerably enhanced due to the fact that the directions 
 of the interacting constituents of the mesogenic unit at the lattice 
 site do not coincide with the orthonormal triad representing the 
 orientation of the molecule, the latter being a necessity to effect 
 tractable reorientations. This requires an intermediate Euler 
 transformation to be performed during the calculation of energy, 
 every time a random step is taken by the molecule. For example 
 typical time for estimating the DoS of this system with size 
 $60 \times 60$ is of the order of 10-12 weeks on a single 
 processor as a serial job.  
  
We attempted to reduce the computing time by adopting a parallel 
computation of the DoS in different sub-segments of the total energy 
range. Adopting the concept suggested in the replica-exchange Monte 
Carlo algorithm \cite{Vogel}, the system energy range of interest is 
divided into \textit{k} equal but significantly overlapping segments. 
The DoS in each segment is determined to the required accuracy 
employing the above procedure, through separate computations 
simultaneously on \textit{k} processors. These components of the Dos 
corresponding to the different energy segments, computed on a 
logarthmic scale, differ from each 
other by an arbitrary constant, which is  specific to each of the 
overlap regions. This allows for synthesizing the total DoS over the 
entire region by stitching them suitably, which can then be used for 
constructing entropic ensemble of the system over the total energy 
range as outlined above. Alternately, one can also determine the 
equilibrium properties of the system corresponding to each energy 
segment from its independently generated entropic (sub)ensemble, 
and finally obtain their variation over the total temperature range 
by similar stitching process for each observable. We find the latter 
procedure to be more practical for our current application. The former 
method determining the DoS over the entire range is however a necessity 
for studying the free energy variations in the space of chosen coordinates.  
For large system sizes entropic sampling based simulations were performed 
with this parallelization choosing \textit{k}= 4, and an overlap of 
70$\%$ between contiguous energy segments. With this choice we have 
large regions of overlap making the stitching process seemless and 
reliable, and we observed nearly 40$\%$ reduction in the computing 
time. We compared this composite data obtained through segmental 
computations with a single window result for consistency, and the 
results matched extremely well. 

MC simulations based on the Metropolis algorithm were also carried 
out for comparison, leading to the collection of B-ensembles. The 
averages are computed  after due equilibration, 
over a  production run comprising of $ 10^{6}$ Monte Carlo lattice sweeps 
(MCS). We found that the derived data on observable variables are 
identical (within errors) from both the procedures. So, in order to 
carry out computations on larger sizes (bigger than 60 $\times$ 60 
lattices) within practical time scales, we opted for Metroplis 
based sampling. 
 
         The simulations were done on square lattices of size 
 $L \times L$ ($L$ = 40, 60, 80, 100), embedded in say laboratory 
 YZ plane, with periodic boundary conditions enforced in the two 
 orthogonal directions. Each lattice site hosts a  symmetric V-shaped 
 molecule with a fixed interarm angle of  $\theta = 112^{\circ}$  
 and the molecules at each  lattice site interact
through the nearest neighbor interaction in eqn.\ref{eqn:1}. For the 
symmetric V-shaped molecule the interarm interaction strengths are equal 
($\epsilon_{AA}=\epsilon_{AB}=\epsilon_{BB}= \epsilon_{BA} =1$). 
The temperature $T$ of the simulation is measured in reduced 
units of $\epsilon_{AA}$. 

   The computed physical observables of interest  are the average 
 energy $<E>$, specific heat $<C_{v}>$, the uniaxial ($R^{2}_{00}$) 
 and biaxial ($R^{2}_{22}$) order parameters of the LC phase \cite{BKL15}. 
Further, we computed topological quantities related to the dominant charge
$1/2$ defects associated with the three order directors. The topological
density \textit{d} is a measure of  abundance of the isolated unbound 
charge $1/2$ defects in the lattice. Its average 
$<d> \ \propto \exp (-E_{0}/T) $ at low temperatures, where 
$<E_{0}>$ is the activation energy required to break a bound defect in 
order to create a pair of oppositely charged defects \cite{Lau89}. 
 Another quantity of interest is the topological 
 order $\mu$ which measures the degree of pairing of defects in 
 a  lattice configuration at that temperature, averaged over the 
 production run. $\mu$ takes values $0 \leq \mu\leq 1$ where $\mu$ = 0 
 denotes the presence of only free defects and $\mu$ = 1 denotes 
 complete pairing. A related derived quantity $\delta = (1-\mu)/2$ 
 is computed and takes values $0\leq \delta\leq 0.5$ \cite{Kunz}. 
 We calculated the topological densities ($d_{x}, d_{y}, d_{z}$) 
and the topological order parameters ($\delta_{x}, \delta_{y}$ and 
$\delta_{z}$) of the x,y,z directors, respectively,  where the 
subscript denotes the defects associated with each of the ordering 
directors. These calculations are described in detail in \cite{BKL20}.
 Pair correlation functions of the spatial variation of reorientational 
 fluctuations, $G(r_{ij})= < P_{2} \ (\cos \theta_{ij})>$, 
 are computed for the three axes (at \textit{L} = 100), denoted as 
 $G_{x}$(r), $G_{y}$(r) and $G_{z}$(r). The above data are computed as 
 a function of temperature in the rangle [0.05, 1.5] with a resolution 
 of 0.005. The correlation functions are computed at 60
 temperatures representatively covering this range. Statistical 
 errors, estimated with the Jack-knife  algorithm \cite{BKL16}, 
 in $E$, $R^{2}_{00}$, $R^{2}_{22}$,
 $ \delta_{(x,y,z)}$ and $d_{x,y,z}$ are typically of the order 
 of 1 in $10^{3}$, while higher moments ($C_{v}$) are relatively 
 less accurate (about 5 in $10^{2}$).

 \section{Results and Discussion} 
\label{.}
While presenting the results for different $L$, we note that data 
at $L$ = 40 and 60 are derived from the WL ensembles and for 
$L$ = 80 and 100 are from B-ensembles. 

\begin{figure}
\centering
\includegraphics[width=0.45\textwidth]{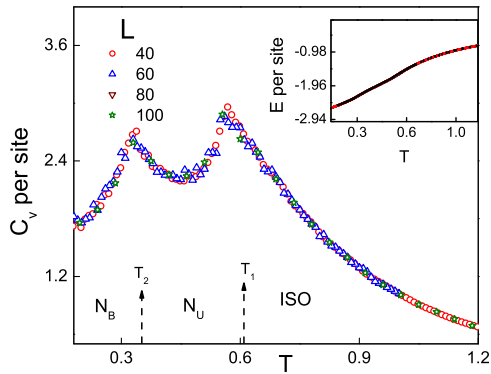}
\caption{(color online) Temperature variation of specific heat 
(per site) at lattice sizes  \textit{L} =  
40, 60, 80, 100. Inset shows the temperature variation of 
 size independent energy per site for different \textit{L}.} 
\label{fig:1}
\end{figure} 
\begin{figure}
\centering
\includegraphics[width=0.45\textwidth]{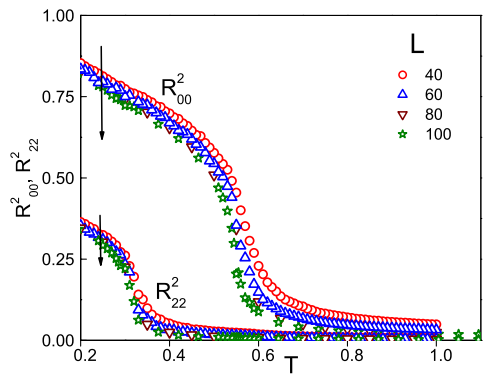}
\caption{(color online) Temperature variation of orientational order 
parameters at lattice sizes  \textit{L} = 40, 
60, 80, 100. (The arrows indicate increase in the size of the system).} 
\label{fig:2}
\end{figure} 

Fig.~\ref{fig:1} depicts the temperature variation of $C_{v}$ (per
site) for system sizes \textit{L} = 40, 60, 80, 100. As the temperature 
is lowered from the isotropic 
phase (at a given system size), the specific heat shows two 
cusps at temperatures $T_{1}$ and $T_{2}$ indicating two phase 
transitions, and are found to be independent of size.
The inset shows the size independence of the energy per site.
In a normal disordering transition, the specific heat peak is a 
measure of the energy fluctuations and the (per site) peak scales 
with the system size. The size independence in this system of the 
peak heights as well as of the profiles of their cusps are early pointers
 to a non-conventional, and possibly topological, 
origin for the occurrence of the transition \cite{CZ88, Kunz}. 
 
 Fig.~\ref{fig:2} shows the temperature
variation of the unixial order parameter $R^{2}_{00}$ and 
the biaxial order parameter $R^{2}_{22}$ for different system sizes. 
At a given size, the sharp increase of uniaxial order near the high 
temperature transition followed by a similar increase in the biaxial 
order at a lower temperature signals changes of symmetry of the LC 
medium to a LC phase of uniaxial symmetry followed by a LC phase of 
biaxial symmetry.  It is seen that the onset temperatures of both 
$R^{2}_{00}$ and $R^{2}_{22}$ shift to lower temperatures and their 
magnitudes decrease as the system size increases. The decrease of low 
temperature order with increase of size is also contrary to 
the expected size variation in conventional order-disorder transition. 
Such size-dependent changes of the orders, which characterize the symmetry 
in the physical space, are established to be specific signatures
 of the topological nature of the two transitions \cite{Botet}.
 Hence these media are to be referred to, more accurately, as LC phases 
 exhibiting corresponding orientational  global symmetries 
 in different temperature regions. We note that they need to be clearly 
distinguished from the conventional uniaxial and biaxial phases, due 
to the qualitative differences they exhibit in their microstructures, 
evidenced by the correlation functions $G(r,T)$. 
For the sake of convenience however, we  refer to these phases 
hereafter as uniaxial and biaxial phases, with this caveat. 
\begin{figure}
\centering
\includegraphics[width=0.45\textwidth]{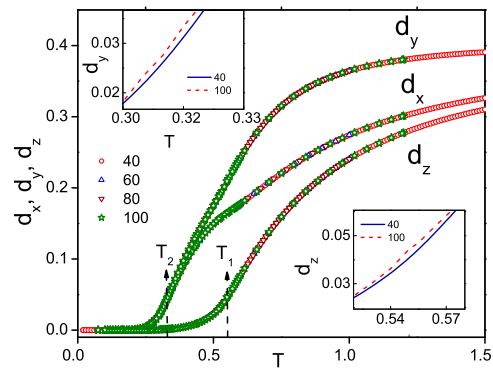}
\caption{(color online) Temperature variation of 
topological densities $d_{x}, d_{y}, d_{z}$ at lattice sizes 
\textit{L} = 40,60, 80, 100. The insets show the slight size 
dependance  for sizes \textit{L} = 40,100 in (a) the biaxial 
phase near $T_{2}$ for $d_{y}$ and  (b) the uniaxial phase
near $T_{1}$ for $d_{z}$.} 
\label{fig:3}
\end{figure} 

The topological parameters (unbound defect density and topological 
order) provide direct evidence to infer about the transition more 
quantitatively. They show sharp changes at the onset of transition, 
clarifying the role of distinct classes of defects that the medium 
hosts. The temperature variation  of the topological density of the 
directors corresponding to x,y,z axes  ($d_{x}, d_{y}, d_{z}$) 
at different lattice sizes (\textit{L} = 40, 60, 80, 100) is depicted in 
Fig.~\ref{fig:3}. At any given size, as the temperature is increased, 
$d_{x}$ and $d_{y}$ increase sharply near the lower temperature transition,
whereas $d_{z}$ increases in the vicinity of the higher temperature
transition. The topological densities are found to be size 
independent over the temperature range but for small 
neighbourhood regions near the two transitions. Their size dependences 
near the two transition regions are depicted in the insets for lattice 
sizes \textit{L} = 40,100. These size dependences of the defect 
densities are reflective of the effect of the system size on the 
process of onset of the BKT-type mechanism of the respective 
transitions.  
\begin{figure}
\centering
\includegraphics[width=0.45\textwidth]{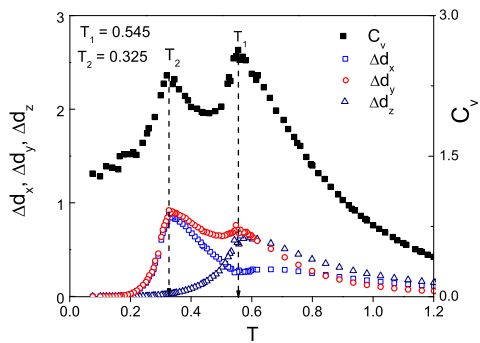}
\caption{(color online) Temperature variation of 
the specific heat $C_{v}$ superposed on the temperature derivatives of 
topological densities ($\Delta d_{x},\Delta d_{y}, \Delta d_{z}$) at 
lattice size \textit{L} =  100.} 
\label{fig:4}
\end{figure} 

Fig.~\ref{fig:4} shows the $C_{v}$ plotted along with derivatives of 
the topological densities ($\Delta d_{\alpha}, \alpha = x,y,z$) as a 
function of the temperature. The cusps of $\Delta d_{\alpha}$ 
($\alpha = x,y,z$) associated with the major and 
minor axes bear striking correlation with the specific heat cusps 
near the transitions from the disordered phase to uniaxial followed 
by biaxial phases, respectively. The derivative cusps are indicative 
of the rate of progression of the proliferation of the unbound defects 
immediately after the onset of the unbinding mechanism of the 
respective transitions. They drive the sharp changes in the specific 
heat variation, indicating absorption of energy due to the unbinding 
process \cite{Berz2, Kosterlitz, Lau89, Holm, Kunz}.  The two  cusps 
are separated in temperature (by $\sim$0.16) which is large enough to 
distinguish the two transitions, but not sufficient to decouple and 
provide data on $G(r,T)$ which is free from pretransitional effects. 
Starting from the low temperature end, we note that the temperature 
variations of $ \Delta d_{x}$ and $\Delta d_{y}$ are practically 
coincident at the onset of the lower transition at $T_{2}$
(Fig.~\ref{fig:3}). They tend to differ significantly with the onset 
of the high temperature transition (in the vicinity of $T_{1}$), 
i.e., at the onset of the proliferation of the major axis defects. 
In the disordered state, $\Delta d_{x}$ and $\Delta d_{z}$  
converge in the high temperature limit, and their limiting 
values are smaller than the saturation value of $\Delta d_{y}$. These 
densities are  measures of mean distances between the corresponding 
unbound defects, and determine relative characteristic lengths 
associated with spatial variations of the corresponding 
oritentational correlations \cite{Kunz}. 

Focussing on the differing profiles of the production rate of unbound 
defects (Fig.~\ref{fig:4}) of the minor axes above $T_{2}$, the rate 
of proliferation of the y-axis defects appears to be only marginally 
affected from its decay profile (with a small cusp at $T_{1}$), and is, 
otherwise, a continuation of its prior path. The growth of the 
unbound defects of the molecular x-axis on the  other hand are 
more profoundly influenced at $T_{1}$. The rapid decay of the 
proliferation rate is temporarily arrested near 
$T_{1}$ before starting to decrease, but at much slower rate.  
Interestingly, above $T_{1}$ the rates of saturation of the z-axis 
and x-axis defect densities coincide asymptotically in the high 
temperature limit.  These subtle qualitative differences in 
$\Delta d_{x}$ and $\Delta d_{y}$ variations near $T_{1}$ are indicative 
of the coupling between the uniaxial and biaxial molecular tensors 
of the interacting sites. For a Hamiltonian with $\gamma$ = 0 for 
example, the Hamiltonian treats the minor axes on equal footing 
and their $\Delta d_{x}$ and $\Delta d_{y}$ profiles were found 
identical at the two transitions through the entire temperature 
range. The defect densities are very subtle but definitive indicators 
of the presence of the $\gamma$ term. The present molecular level 
model thus corresponds to a general biquadratic Hamiltonian in its 
phenemenological description with two independent model parameters 
to account for the biaxial interactions in the system.  
 \begin{figure}
\centering
\includegraphics[width=0.45\textwidth]{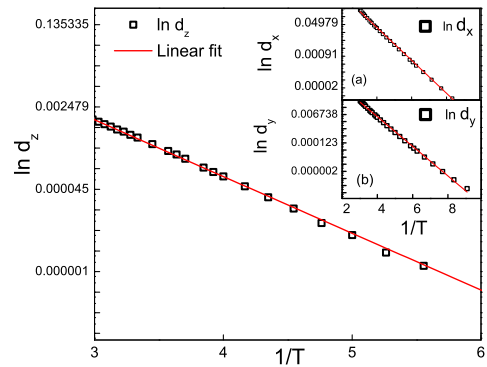}
\caption{(color online) Plot of natural logarithm of 
defect density $d_{z}$ versus the inverse temperature at lattice 
size \textit{L} = 100. A straight line 
fit to the curve gives an estimate of activation energy for z-defects.
Insets show similar plots for (a) $d_{x}$  and (b) $d_{y}$ } 
\label{fig:5}
\end{figure}  

The onset of the topological transition is initiated by a thermal 
activation process and we estimate this energy for each category 
of defects by fitting their data on initial growth of the unbound 
defect density to the Arrhenius equation \cite{Lau89}. Fig.~\ref{fig:5} 
depicts variation of the three defect densities (on a log scale) with 
respect to the inverse of temperature. The data fit very satisfactorily 
to straight lines in each case, and the magnitudes of the slopes are 
measures of the activation energies associated with the unbinding 
mechanisms of the respective transitions. This energy value for the 
transition at $T_{1}$ due to z-axis defects is $E_{z}$ = 2.75$\pm$ 0.023 
(Fig.~\ref{fig:5}). The unbound defects of x-and y-axes are activated 
with identical energy and the corresponding values are : 
$E_{x}$ = $E_{y}$ = 2.08$\pm$ 0.017 (insets of Fig.~\ref{fig:5}).
\begin{figure}
\centering
\includegraphics[width=0.45\textwidth]{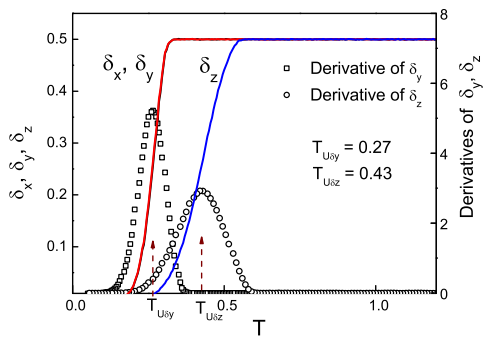}
\caption{(color online) Temperature variation of topological order 
parameters  $\delta_{x}, \delta_{y}, \delta_{z}$at  \textit{L} = 100. 
The dotted curves are the temperature 
derivatives of $\delta_{z}$ and $\delta_{y}$ having peaks at 
at $T_{U\delta z}$ and $T_{U\delta y}$}. 
\label{fig:6}
\end{figure} 
\begin{figure}
\centering
\includegraphics[width=0.45\textwidth]{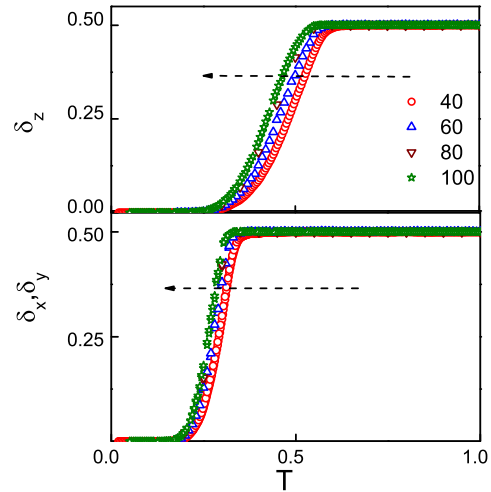}
\caption{(color online) Temperature variation of topological order 
parameters $\delta_{x}, \delta_{y}, \delta_{z}$ at lattice sizes  
\textit{L} = 40,60, 80, 100 (The arrows indicate the increase in 
system size).} 
\label{fig:7}
\end{figure}

Fig.~\ref{fig:6} depicts the temperature variation of the  
topological variables of the x,y,z axes ($\delta_{x}, \delta_{y}, 
\delta_{z}$) related to the corresponding topological order and 
their temperature derivatives, at \textit{L} = 100. As the system 
is cooled starting from the isotropic phase,  $\delta_{z}$ sharply 
decreases from a constant value of 0.5, at the onset of the high 
temperature transition, and  in the completely (topologically) 
ordered state (of z-axes defects), its value is zero. We note that 
the corresponding parameters of x and y axes, $\delta_{x},\delta_{y}$,  
are unaffected by this transition. They  
show however  similar changes with temperature at the second transition. 
These sharp changes in $\delta$ variables are definitive markers 
identifying the category of defects associated with the particular 
transition. The inflexion point of their decay identifies the 
transition temperature where the unbinding mechanism is formally 
initiated.  We denote the corresponding transition temperatures for 
this system size (\textit{L} = 100) as $T_{U\delta z}(L)$ = 0.43
($\pm$ 0.005) and $T_{U\delta x}(L) = T_{U\delta y}(L) $ = 0.27 
($\pm$ 0.005), indicated in Fig.~\ref{fig:6}. The size dependence of 
the topological order profiles is shown in Fig.~\ref{fig:7}.
The lowering of transition temperatures with increase in size is in 
accord with similar size variation of the onset of orientational 
order parameters (Fig.~\ref{fig:2}).  
\begin{figure}
\centering{
\subfigure[]{\includegraphics[width=0.45\textwidth]{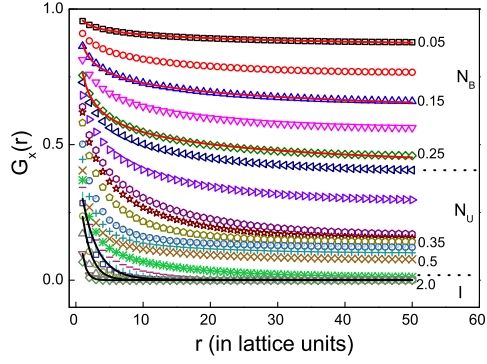}
\label{fig:8a}}
\subfigure[]{\includegraphics[width=0.45\textwidth]{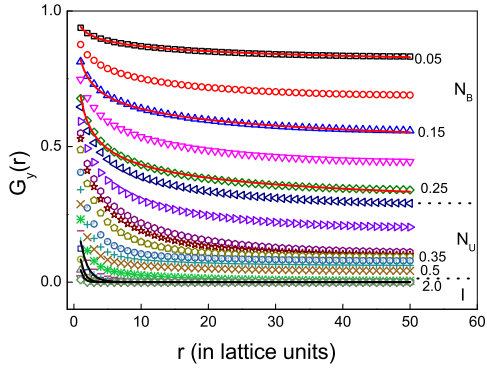}
\label{fig:8b}}
\subfigure[]{\includegraphics[width=0.45\textwidth]{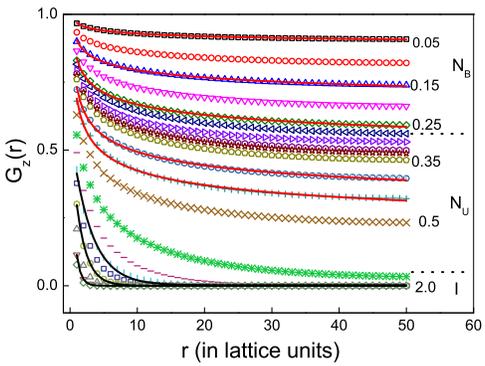}
\label{fig:8c}}
}
\caption{(color online) Spatial variation of the correlation 
functions $G(r)$ of the x,y,z directors (\textit{L} = 100) at chosen 
temperatures in the isotropic, uniaxial and biaxial phases. $G(r)$ fits
 to exponential decays in the isotropic phase (Black solid line fit 
 superposed on the data), power law decays (red dotted line fit 
 superposed on the data) in the biaxial phase for all axes whereas 
 in the uniaxial phase only $G_{z}(r)$ exhibits power law decay.   } 
\label{fig:8}
\end{figure}

Correlation functions $G(r,T)$ of the x,y,z directors
 at \textit{L}=100 at certain chosen temperatures 
 (out of the data collected at 60 temperatures) are shown in 
 Fig.~\ref{fig:8}, covering all the three phases of the model. 
 These follow definite analytical decay in the isotropic (above $T_{1}$) 
 and biaxial symmetry (below $T_{2}$) phases. These fit very well to 
 exponential decays with a different correlation length for each 
 director, as $G(r,T)= A exp \ (-r/\xi_{\alpha}(T))+ C$ , 
 ($\alpha$ = x, y, z) in the isotropic phase. In the phase with 
 biaxial symmetry, they follow power law decays, each with its 
 own exponent, as $G(r,T) \approx r^{-\eta_{\alpha}(T)}$, 
 ($\alpha$ = x, y, z). For temperatures in the range $0.3\leq T \leq 0.555$ 
 (uniaxial symmetry) correlation functions of x and y directors 
 do not fit satisfactorily to either a power law or 
an exponential decay, whereas the correlation function of the 
z director fits very well to a power law.  
\begin{figure}
\centering{
\subfigure[]{\includegraphics[width=0.45\textwidth]{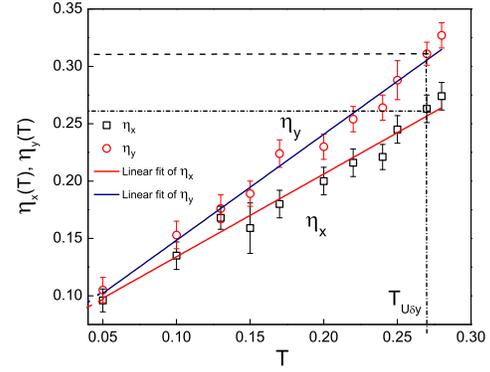}
\label{fig:9a}}
\subfigure[]{\includegraphics[width=0.45\textwidth]{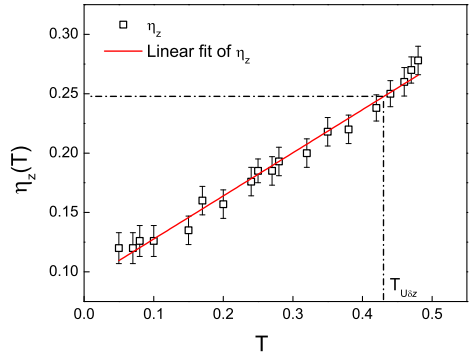}
\label{fig:9b}}
}
\caption{(color online)Temperature variation of the power law exponents 
$\eta_{x}$(T), $\eta_{y}$(T) in the biaxial phase and $\eta_{z}$(T) 
in uniaxial and biaxial phases. } 
\label{fig:9}
\end{figure}

The temperature variation of the power law exponents in the 
uniaxial and biaxial phases  is depicted in Fig.~\ref{fig:9}.  
$\eta_{x}$(T) and $\eta_{y}$(T) 
(Fig.~\ref{fig:9a}) decrease linearly with temperature as 
$T\rightarrow 0$ in the biaxial phase with different slopes. They 
are expected to vanish at $T=0$ in large enough samples approximating 
the thermodynamic result. In the present case (\textit{L}=100) they 
tend to a non-vanishing value in this temperature limit, which is an 
artefact of the finite size of the system. We note that
$\eta_{x}$(T)= 0.26 and $\eta_{y}$(T) = 0.31 at the unbinding
temperatures $T_{U\delta x} = T_{U\delta y}$ = 0.27, reasonably 
close to the mean-field expected value of 0.25 (for 2D-XY and 
planar LL model \cite{Botet}). Fig.~\ref{fig:9b} depicts variation 
of  $\eta_{z}$(T) over the temperature range covering
 uniaxial and biaxial symmetric phases, with a value $\eta_{z}$ = 0.25  
at  $T_{U\delta z}$ = 0.43. 
\begin{figure}
\centering
\includegraphics[width=0.45\textwidth]{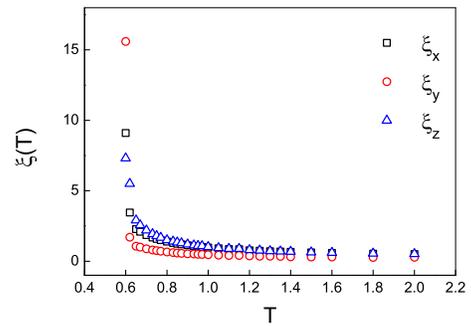}
\caption{(color online) Temperature variation of the correlation 
lengths $\xi_{x}(T)$, $\xi_{y}(T)$ and $\xi_{z}(T)$ of x, y and
z-defects  at lattice size \textit{L} = 100.} 
\label{fig:10}
\end{figure}

Temperature dependences of the  three correlation lengths in the 
isotropic phase, $\xi_{x}(T)$, $\xi_{y}(T)$ and $\xi_{z}(T)$, are 
depicted in Fig.~\ref{fig:10}. We observe the expected correlation of 
the observed magnitudes of the unbound defect densities associated 
with the three directors above the high temperature transition 
(Fig.~\ref{fig:3}), with their correlation lengths. The y-axis 
defects have a lower correlation length $\xi_{y}(T)$  value 
(until their divergences set in) than the 
others. The x- and z-director defects have comparable values with 
$\xi_{z}(T)$ marginally becomes greater than $\xi_{x}(T)$ particularly 
as the transition point is reached from above. This is in accord with 
the expectation from their defect densities, and establishes clearly 
the origin of length scales in this phase.    
 
 For the transition observed at $T_{1}$ the critical behaviour of 
 $\xi_{z}(T)$ alone is obviously relevant. The differing divergences 
of the other two correlation lengths are reflective of the differential
perturbations that the corresponding unbound defect densities suffer 
due to this transition. These reflect the cross coupling of the uniaxial 
and biaxial tensors in the phenomenological description. The relevant 
correlation length $\xi_{z}(T)$ should show essential 
 divergence near a topological transition, unlike a simple divergence 
 as is the case with conventional transitions. The mean field expression 
 describing the divergence near a topological transition arising from 
 the sharp disappearance of the unbound defects is given by 
$\xi(T)\approx \exp \ \left[\frac{D}{(T-T_{U\delta})^{\nu}}\right]$
\cite{Kosterlitz, Kenna, Kawamura, BKLPRL, BKL20}, where $T_{U\delta}$ 
is the unbinding temperature determined earlier and $\nu$ is the 
associated critical index. 
\begin{figure}
\centering
\includegraphics[width=0.45\textwidth]{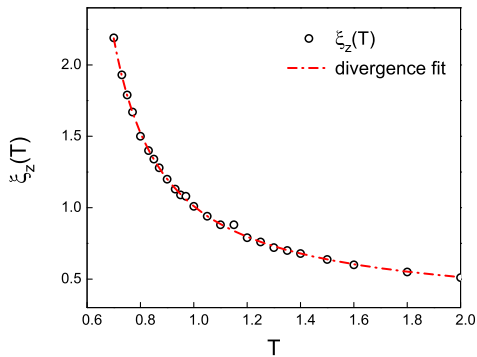}
\caption{(color online) Temperature variation of the correlation 
length $\xi_{z}(T)$ of z-defects  at lattice size  
\textit{L} = 100. The dash-dotted line is the divergence fit of 
$\xi_{z}(T)$ as indicated in the text.} 
\label{fig:11}
\end{figure}
The fit using the predetermined value $T_{U\delta z}$ = 0.43,
shown in Fig.~\ref{fig:11}, yields $\nu$ = 0.5 
($\pm$ 0.025), which compares very well with the mean field value. 
The close proximity of the two transitions, coupled with the evidence
of cross coupling interaction, prohibits single length scales for 
correlation functions associated with the minor axes in the 
uniaxial medium, and hence no useful information 
could be inferred about the critical properties of the second transition.
\section{Conclusions}
The present study based on Monte Carlo simulations establishes clearly 
two topological transitions starting from a disordered phase in this 
two-dimensional V-shaped bent-core model. The first high temperature 
transition imparts uniaxial symmetry to the system and the second 
transition condenses the system with biaxial symmetry. The onset of 
the orientational orders at different sizes is correlated with the 
corresponding specific heat data. The size dependences of the 
orientational order(s) in the respective phases indicate underlying 
topological defect mediated mechanism for these transitions.
Temperature variations of topological properties, viz. unbound defect 
density associated with the order directors as well as their topological 
order parameters, provide direct confirmation of the topological 
origin of the transitions, besides determining the transition 
temperatures with a high degree of accuracy. Analysis of $G(r,T)$ data 
provides quantitative information on the critical indices as well as 
activation energies associated with the unbinding mechansims 
associated with the two transitions, besides demonstrating qualitative 
changes in the microscopic structures of the medium with temperature. 
In particular, in the format of a phenomenological description of the 
Hamiltonian of this model in terms 
of general biquadratic interactions among uniaxial and biaxial 
molecular tensors, the data on topological parameters indicate the 
presence of an appreciable degree of cross coupling between the 
neighbouring lattice sites.  The low temperature biaxial phase with 
the observed QLRO (with respect to the molecular axes) is characterized 
by a line of critical points below the second transition temperature.   

 \section{Acknowledgments}
We acknowledge the computational support from the Centre for  Modelling
 Simulation and Design (CMSD) and the School of Computer and 
Information Sciences (DST PURSE - II Grant) at the University of
 Hyderabad. BKL acknowledges financial support from Department of
 Science and Technology, Government of India  vide grant ref No: 
 SR/WOS-A/PM-2/2016 (WSS) to carry out this work. SD acknowledges 
 grant from SERB (Ref. No. CRG/2019/000425)

\end{document}